\documentclass[preprint,10pt]{elsarticle}

\usepackage{textcomp,booktabs}
\usepackage[usenames,dvipsnames]{color}
\usepackage{colortbl}
\definecolor{mygray}{gray}{.9}
\definecolor{mypink}{rgb}{.99,.91,.95}
\definecolor{mycyan}{cmyk}{.3,0,0,0}
\usepackage{amssymb}
\usepackage{graphicx}
\usepackage{booktabs}
\usepackage{tabularx}
\usepackage{array}
\usepackage{lscape}
\usepackage{longtable}
\usepackage{multirow}
\usepackage{amsmath}
\usepackage{colortbl}
\usepackage{epsfig}
\usepackage{epsf}
\usepackage{subfigure}
\usepackage{rotating}
\usepackage{pdflscape}
\usepackage{lscape}
\usepackage{indentfirst}
\usepackage{wrapfig}
\usepackage{enumerate}
\usepackage{amsfonts,amssymb}
\usepackage{dsfont}
\usepackage{CJK}
\usepackage[linesnumbered,ruled]{algorithm2e}
\usepackage[amsmath,thmmarks]{ntheorem}
\biboptions{numbers,sort&compress}

\newcommand{\PreserveBackslash}[1]{\let\temp=\\#1\let\\=\temp}
\newcolumntype{C}[1]{>{\PreserveBackslash\centering}p{#1}}
\newcolumntype{R}[1]{>{\PreserveBackslash\raggedleft}p{#1}}
\newcolumntype{L}[1]{>{\PreserveBackslash\raggedright}p{#1}}
\newtheorem{definition}{Definition}[section]

\begin{document}

\begin{frontmatter}

\title{A generalized intelligent quality-based approach for fusing multi-source information}

\author[address1]{Fuyuan Xiao\corref{label1}}
\ead{xiaofuyaun@swu.edu.cn}
\address[address1]{School of Computer and Information Science, Southwest University, Chongqing, 400715, China}
\cortext[label1]{Corresponding author at: School of Computer and Information Science, Southwest University, No.2 Tiansheng Road, BeiBei District, Chongqing, 400715, China. 
}

\begin{abstract}
In this paper, we propose a generalized intelligent quality-based approach for fusing multi-source information.
The goal of the proposed approach intends to fuse the multi-complex-valued distribution information while maintaining a high quality of the fused result by considering the usage of credible information sources.
\end{abstract}


\end{frontmatter}

\section{Introduction}\label{Introduction}
In this paper, inspired by Yager and Petry's~\cite{yager2016intelligent}, a complex-valued intelligent quality-based methodology is proposed for multi-source information fusion.
To be specific, a vector representation of complex-valued distribution is first defined.
Some new concepts used in complex-valued distributions are presented, including the compatibility degree and conflict degree between complex-valued distributions.
Based on that, the intelligent information quality measure of complex-valued distributions is devised by leveraging the concept of Gini entropy~\cite{gini1955variabilita}.

\newtheorem{Remark}{Remark}
\newtheorem{Property}{Property}
\newtheorem{Theorem}{Theorem}
\newtheorem*{Proof}{Proof}
\newtheorem{exmp}{Example}

\section{Information quality measure of complex-valued distribution}\label{Informationquality}

\begin{definition}(Complex-valued distribution vector)\label{def_CvD}

Let $\mathbb{C}_k$ be a complex-valued distribution (CvD) vector on the space $\Psi = \{\psi_1, \psi_2, \ldots, \psi_j, \ldots, \psi_n\}$, denoted by
\begin{equation}
\mathbb{C}_k = [c_{k1}, c_{k2}, \ldots, c_{kj}, \ldots, c_{kn}],
\end{equation}
where $c_{kj}$ represents the complex value of the occurrence of $\psi_j$, denoted by
\begin{equation}\label{eq_complexnumber}
c_{kj} = x_{kj} + y_{kj}i,
\end{equation}
where $x_{kj}$ and $y_{kj}$ are real numbers and $i$ is the imaginary unit, satisfying $i^2 = -1$.

In Eq.~(\ref{eq_complexnumber}), for $c_{kj}$, it satisfies the conditions:
\begin{equation}
\begin{aligned}
&x_{kj} \geq 0, \\
&\sqrt{x_{kj}^2+y_{kj}^2} \in [0, 1], \\
&\sum^n_{j=1} c_{kj} = 1.
\end{aligned}
\end{equation}

\end{definition}

\begin{definition}(The inner product between CvDs)

Let $\mathbb{C}_k$ and $\mathbb{C}_h$ be two complex-valued distribution vectors on the space $\Psi$.
The inner or dot product of $\mathbb{C}_k$ and $\mathbb{C}_h$ is defined by
\begin{equation}
\langle\mathbb{C}_k, \mathbb{C}_h\rangle = 
\mathbb{C}_k \cdot \mathbb{C}_h
= \sum^n_{j=1} c_{kj} \overline{c}_{hj}.
\end{equation}
\end{definition}

Then, the norm of the complex-valued distribution vector is defined as
\begin{equation}\label{eq_normCvDV}
\|\mathbb{C}_k\|
= \sqrt{\langle\mathbb{C}_k, \mathbb{C}_k\rangle}.
\end{equation}
\begin{definition}(Cosine of the angle between CvDs)

The Cosine of the angle between complex-valued distribution vectors $\mathbb{C}_k$ and $\mathbb{C}_h$, denoted as $\cos \Theta(\mathbb{C}_k, \mathbb{C}_h)$ is defined by
\begin{equation}\label{eq_Cosineangle}
\cos \Theta(\mathbb{C}_k, \mathbb{C}_h)
= \frac{\langle\mathbb{C}_k, \mathbb{C}_h\rangle + \langle\mathbb{C}_h, \mathbb{C}_k\rangle}{2\|\mathbb{C}_k\| \|\mathbb{C}_h\|}.
\end{equation}
\end{definition}

\begin{definition}(The compatibility degree between CvDs)\label{def_comCvDVs}

Let $\mathbb{C}_k$ and $\mathbb{C}_h$ be two complex-valued distribution vectors on the space $\Psi$.
The compatibility degree between CvDs $\mathbb{C}_k$ and $\mathbb{C}_h$, denoted as $Com(\mathbb{C}_k, \mathbb{C}_h)$ is defined as
\begin{equation}\label{eq_comCvDVs}
Com(\mathbb{C}_k, \mathbb{C}_h)
= \frac{|\langle\mathbb{C}_k, \mathbb{C}_h\rangle + \langle\mathbb{C}_h, \mathbb{C}_k\rangle|}{2\|\mathbb{C}_k\| \|\mathbb{C}_h\|}.
\end{equation}
\end{definition}

The properties of $Com(\mathbb{C}_k, \mathbb{C}_h)$ are:
\begin{enumerate}[(1)]
\item
\textbf{Symmetry:} $Com(\mathbb{C}_k, \mathbb{C}_h)=Com(\mathbb{C}_h, \mathbb{C}_k)$;

\item
\textbf{Boundedness:} $Com(\mathbb{C}_k, \mathbb{C}_h) \in [0, 1]$;

\item
\textbf{Non-degeneracy:} $Com(\mathbb{C}_k, \mathbb{C}_h) = 1$ if and only if $\mathbb{C}_k = \mathbb{C}_h$;

\item
\textbf{Orthogonality:} $Com(\mathbb{C}_k, \mathbb{C}_h) = 0$ if and only if for $k$ we have $c_{kj} \neq 0$ and $c_{hj} = 0$, and for $h$ we have $c_{hj} \neq 0$ and $c_{kj} = 0$, so that for each we obtain $c_{kj}c_{hj}=0$.

\end{enumerate}

\begin{definition}(The conflict degree between CvDs)

Let $\mathbb{C}_k$ and $\mathbb{C}_h$ be two complex-valued distribution vectors on the space $\Psi$.
The conflict degree between CvDs $\mathbb{C}_k$ and $\mathbb{C}_h$, denoted as $Con(\mathbb{C}_k, \mathbb{C}_h)$ is defined as
\begin{equation}\label{eq_conCvDVs}
Con(\mathbb{C}_k, \mathbb{C}_h)
= 1-Com(\mathbb{C}_k, \mathbb{C}_h).
\end{equation}
\end{definition}

\begin{definition}(Information quality for CvD)\label{def_CvDIQ}

Let $\mathbb{C}_k$ be a complex-valued distribution vector on the space $\Psi$.
The information quality of $\mathbb{C}_k$, denoted as $IQ(\mathbb{C}_k)$ is defined as
\begin{equation}\label{eq_conCvDVs}
IQ(\mathbb{C}_k)
= \|\mathbb{C}_k\|^2.
\end{equation}

\end{definition}

\begin{definition}(Information quality for multi-CvDs)

The complex-valued information quality of $Agg$ for multi-CvDs is defined as:
\begin{equation}\label{eq_complexIQ}
\|Agg\|^2 =
\frac{1}{r^2} \left[\sum^r_{k=1} \|\mathbb{C}_k\|^2 + 2 \sum^{r-1}_{k=1} \sum^r_{h=k+1} \frac{\mathbb{C}_k \cdot \mathbb{C}_h + \mathbb{C}_h \cdot \mathbb{C}_k}{2}\right].
\end{equation}
\end{definition}

In particular, when complex-valued distributions to probability distributions become probability distributions, the $\|Agg\|^2$ degrades into $\|R\|^2$ of Yager and Petry's.

\section{Conclusions}\label{Conclusion}
In this paper, we proposed an intelligent quality-based methodology.
This intelligent quality-based methodology can be used for multi-source information fusion of complex-valued distributions.
The main contribution is that the complex-valued intelligent quality-based methodology was a generalization of Yager and Petry's~\cite{yager2016intelligent}.
In particular, when the complex-valued distributions became probability distributions, the proposed intelligent quality-based methodology degraded into Yager and Petry's.



\clearpage
\normalsize
\bibliographystyle{elsarticle-num}

\end{document}